\documentclass[a4paper,11pt]{article}
\usepackage{pos}
\usepackage{algorithm}
\usepackage{algpseudocode}

\DeclareMathAlphabet{\bbold}{U}{bbold}{m}{n}
\newcommand{\id}{\ensuremath{\bbold{1}}} % For double-struck 1

\title{Optimized solver algorithms for computing $\mathrm{Tr}(A^{-1})$}
\author*[a]{Benjamin Luke}
\author[a]{Sudip Shiwakoti}
\author[a]{Shayan Nadeem}
\author[b]{Ronald Morgan}
\author[a]{Walter Wilcox}

\affiliation[a]{Department of Physics and Astronomy, Baylor University,\\
  Waco, Texas, 76798, USA}
\affiliation[b]{Department of Mathematics, Baylor University,\\
  Waco, Texas, 76798, USA}

\emailAdd{Ben\_Luke1@baylor.edu}
\emailAdd{Sudip\_Shiwakoti1@baylor.edu}
\emailAdd{Shayan\_Nadeem1@baylor.edu}
\emailAdd{Ronald\_Morgan@baylor.edu}
\emailAdd{Walter\_Wilcox@baylor.edu}

\abstract{Estimating the trace of the inverse of a large quark matrix is a longstanding problem in lattice QCD. The standard approach is Hutchinson's method, which uses random noise vectors to estimate the trace to a desired statistical tolerance. Each noise vector $b$ requires solving a system of linear equations $Ax = b$. In this work, we introduce a new class of Krylov solvers for this task, called Twin methods. Twin methods generate the Krylov polynomial once then reuse it on all other right-hand sides. This reduces the number of basic linear algebra operations in the algorithm and therefore lowers the trace computation time. We apply the Twin approach to BiCG and BiCGStab and test them on lattice QCD matrices of varying size and in a range of computing contexts. We also present a more economical estimation of $\mathrm{Tr}(A^{-1})$ which implements a polynomial approximation of $A^{-1}$ and a singular value deflation. In this scheme, the solvers are applied to only a few noise vectors, and the remainder are treated by the polynomial and the deflation. In our tests, the polynomial approximation combined with the deflation yield a 17-fold speedup in the time required to compute the trace over standard Hutchinson's method.} 

\FullConference{The 43rd International Symposium on Lattice Field Theory (Lattice 2026)\\
July 26 to August 1, 2026\\
University of Maryland, College Park, USA\\}

\begin{document}
\maketitle

\section{Introduction}
A longstanding problem in lattice QCD is computing the trace of the inverse Wilson-Dirac operator, $\mathrm{Tr}(A^{-1})$, which corresponds to the fermionic contraction of the scalar density operator, $\bar{\psi}\psi$, on the lattice. Because computing $A^{-1}$ directly is prohibitively expensive, the trace is estimated stochastically. The standard approach is Hutchinson's method \cite{HutchTrace},
\begin{equation}
\mathrm{Tr}(A^{-1}) \approx \frac{1}{N} \sum_{i = 1}^N b_i^\dagger A^{-1} b_i = \frac{1}{N} \sum_{i = 1}^N b_i^\dagger x_i,
\end{equation}
where the $b_i$ are isotropic random noise vectors and $x_i = A^{-1}b_i$ is the solution to the system of linear equations $Ax_i = b_i$. In practice, $N$ is not fixed but is determined by the number of noise vectors required to reach a desired standard-error tolerance. %The cost of the trace therefore is determined by the cost of $N$ linear solves.

There are many algorithms for solving large systems of linear equations, including BiCG and BiCGStab in addition to more sophisticated methods. In Sec.~\ref{sec:Algorithms} we introduce a new class of Krylov solvers for efficiently solving multiple RHSs, which we call ``Twin'' methods. A Twin method solves the first RHS using a standard Krylov solver then reuses the Krylov polynomial on all subsequent RHSs. This approach significantly reduces the number of basic linear algebra operations that contribute to the cost of the problem.

This paper applies the Twin method to BiCG and BiCGStab and compares them with standard BiCGStab and BiCG-QCD (BiCG exploiting the QCD $\gamma^5$ symmetry). In Sec.~\ref{sec:TwinTesting}, we present tests on various QCD matrices and interpret the results. Sec.~\ref{sec:PSandSVD} discusses a more elaborate evaluation of the trace using a polynomial approximation of $A^{-1}$ and a singular value deflation, which reduces the computational expense.

\section{Algorithms} \label{sec:Algorithms}
The algorithms we use in this work are BiCGStab \cite{BiCGStab}, BiCG-QCD \cite{FROMMER_1995}, Twin BiCG, and Twin BiCGStab. Table~\ref{tbl:methodsSummary} summarizes the basic features of the first three methods. We first present the theoretical basis for Twin Krylov methods (Sec.~\ref{sec:TwinBiCG}) and then discuss their benefits over independent, sequential Krylov solvers (Sec.~\ref{sec:TwinBenefits}).

\begin{table}[h] \centering
\begin{tabular}{l|p{2.4cm}p{2.4cm}p{2.4cm}p{2.6cm}} \hline\hline
Method                  & BiCGStab                   & BiCG-QCD                                     & Twin BiCG             & Twin BiCGStab            \\ \hline
MVPs/iteration      & 2                          & 1                                            & 1*             & 2*                \\
DPs/iteration       & 2                          & 1                                            & 0*             & 0*                       \\
Multiple solves		 & independent        & independent          & reuse $\pi$ polynomial      & reuse $\pi$ polynomial           \\\hline\hline
\end{tabular}
\caption{Solver characteristics. Asterisks indicate costs after the base solve.\label{tbl:methodsSummary}}
\end{table}

\subsection{Twin Krylov methods} \label{sec:TwinBiCG}
To motivate Twin Krylov methods, we first review the convergence of a general Krylov method. To begin, note that the approximate solution $\hat{x}$ to $Ax = b$ is extracted from the Krylov subspace,
\begin{equation}
\hat{x} \in \mathcal{K}^{m}(b,A) = \text{span}\{b,Ab,A^2b,\dots,A^{m-1}b\}.
\end{equation}
Thus $\hat{x} = p(A)b$ for some polynomial $p(A)$ of degree $m-1$ or less, and the residual is 
\begin{equation}
r = b - A\hat{x} = \pi(A)b,
\end{equation}
where $\pi(A) = I - Ap(A)$ is the residual polynomial. Let $\{\lambda_i, z_i\}$ be the orthonormal eigenbasis of the $n\times n$ matrix $A$ and expand $b = \sum_i c_i z_i$. Then 
\begin{align}
\pi(A)b &= \pi(A)(c_1 z_1 + \cdots + c_n z_n) \\
&= c_1\pi(\lambda_1)z_1 + \cdots + c_n \pi(\lambda_n)z_n.
\end{align}
The residual is therefore small when $\pi(\lambda_i)$ is small along the whole spectrum of $A$, at least for the eigenvectors that $b$ has substantial weight on.

The Twin method for multiple RHSs relies on the following observation: $\pi$ is determined by the part of the spectrum of $A$ where the first RHS has weight, and a $\pi$ that is small along the eigencomponents of one RHS will also be small along the shared eigencomponents of any other RHS. Since the random vectors are isotropic, $\mathbb{E}[bb^\dagger] = \id$, and the eigenbasis is orthonormal, then the eigencomponents are uniformly distributed for each noise, $\mathbb{E}[\|c_i\|^2] = 1$. Hence, even though $\pi$ is generated using a particular noise, it does not statistically favor that noise and may be applied unbiasedly to other RHSs.\footnote{One may object that since the Wilson-Dirac matrix is non-normal, an orthonormal eigenbasis is not guaranteed and the argument above is not general; as presented, it is merely heuristic. However, one may show an even more general result: the expectation of the residual norm is independent of the noise, $\mathbb{E}[\|\pi(A)b\|^2] = \|\pi\|^2_F$, where $\|\cdot\|_F^2$ is the Frobenius norm. Hence the methodology still holds for QCD matrices.}

In practice, this amounts to saving the recurrence coefficients generated on the first solve and reusing them on the next solves. Alg.~\ref{alg:TwinBiCG} shows the Twin approach applied to BiCG, which we call Twin BiCG.

\begin{algorithm}
\begin{algorithmic}[1]
\State Solve the first RHS using standard BiCG and save the parameters $\{\alpha_i\}$ and $\{\beta_i\}$. For all subsequent RHSs:
\State Compute $r_0 = b-Ax_0$. Choose $r_0'$ such that $r_0'^\dagger r_0 \neq 0$.
\State Set, $p_0 = r_0$, $p_0' = r_0'$.
\For{$j = 0,1,\dots m$}
	%\State $\alpha_j = r_j'^{\dagger}r_j/p_j'^{\dagger}(Ap_j)$
	\State $x_{j+1} = x_j + \alpha_j p_j$
	\State $r_{j+1} = r_j - \alpha_j A p_j$
	%\State $r'_{j+1} = r'_j - \alpha_j A^\dagger p'_j$
	%\State $\beta_j = r'^\dagger_{j+1} r_{j+1}/r'^\dagger_{j} r_{j}$
	\State $p_{j+1} = r_{j+1} + \beta_j p_j$
	%\State $p'_{j+1} = r'_{j+1} + \beta_j p'_j$
\EndFor
\end{algorithmic}
\caption{Twin BiCG (for multiple RHSs) \label{alg:TwinBiCG}}
\end{algorithm}

\subsection{Benefits of Twin algorithms} \label{sec:TwinBenefits}
Beyond the first solve, Twin BiCG halves the number of matrix-vector products (MVPs) relative to standard BiCG and both Twin BiCG and Twin BiCGStab eliminate dot products entirely. The latter matters for large scale problems on distributed memory where the relative cost of a dot product is very high due to the overhead cost to transfer the information. %Both savings apply to the basic linear algebra operations common to all solvers.

Twin methods also adapt easily to a batched solve context where multiple RHSs are solved simultaneously as $AX = B$, where $B = [b_1 \cdots b_m]$ for $m$ RHSs, and $X$ contains the corresponding solutions. Batching replaces matrix-vector operations with matrix-matrix operations, which reuse data more efficiently. Because the coefficients are shared, each column of $B$ requires the same number of iterations and therefore the noise vectors converge together. This simplifies the implementation compared to other batched Krylov solvers which require a different number of iterations for each column in $B$. 

Furthermore, Twin methods can prevent convergence failures in systems that are prone to them. Suppose standard BiCGStab fails on one in four noise vectors, meaning no usable $\pi$ is generated. Twin BiCGStab avoids this since the parameters are saved from the first successful RHS and reused on all others, including those that would otherwise fail under standard BiCGStab. In Sec. \ref{sec:TwinTesting} we will see several instances where standard BiCGStab experiences convergence failures on noise vectors for which Twin BiCGStab does not.

\section{Hutchinson's Method Tests} \label{sec:TwinTesting}

Tests were performed on QCD matrices from three ensembles. In Sec.~\ref{sec:singleThread}, matrices labeled ``H'' come from a $12^3\times 16$ quenched lattice, with Wilson fermions at the critical hopping parameter $\kappa_c = 0.157$. In Secs.~\ref{sec:Amats} and \ref{sec:Bmats}, we use gauge configurations from the MILC Collaboration's $N_f = 2+1+1$ Highly Improved Staggered Quark (HISQ) program \cite{PhysRevD.87.054505}, generated with a one-loop Symanzik-improved gauge action and HISQ dynamical sea quarks. Ensemble ``A'' is $16^3\times 48$ with light quark mass $m_{l,A} = 16.9$ MeV, and ensemble ``B'' is $24^3\times 64$ with $m_{l,B} = 8.26$ MeV. Wilson fermion matrices are built on these configurations with hopping parameters corresponding to the physical pion mass: $\kappa_A = 0.1708$ and $\kappa_B = 0.1657$. The GPU tests in Sec.~\ref{sec:Amats} onward were run on the Delta cluster using NVIDIA A100 GPUs.

\subsection{Single thread benchmark comparison} \label{sec:singleThread}
We begin by comparing the computational cost of Twin BiCG to BiCGStab using minimal computer resources. The tests were run on the H matrices and utilized a single Delta CPU core and 128 Gb of RAM. In MATLAB, the number of computational threads was restricted to one since multiple threads are used by default. This isolates the benefits of algorithm design from those of hardware parallelism.

\newcommand{\figWidth}{4.7} 

\begin{figure}[h] 
   \centering
   \includegraphics[width = \figWidth in]{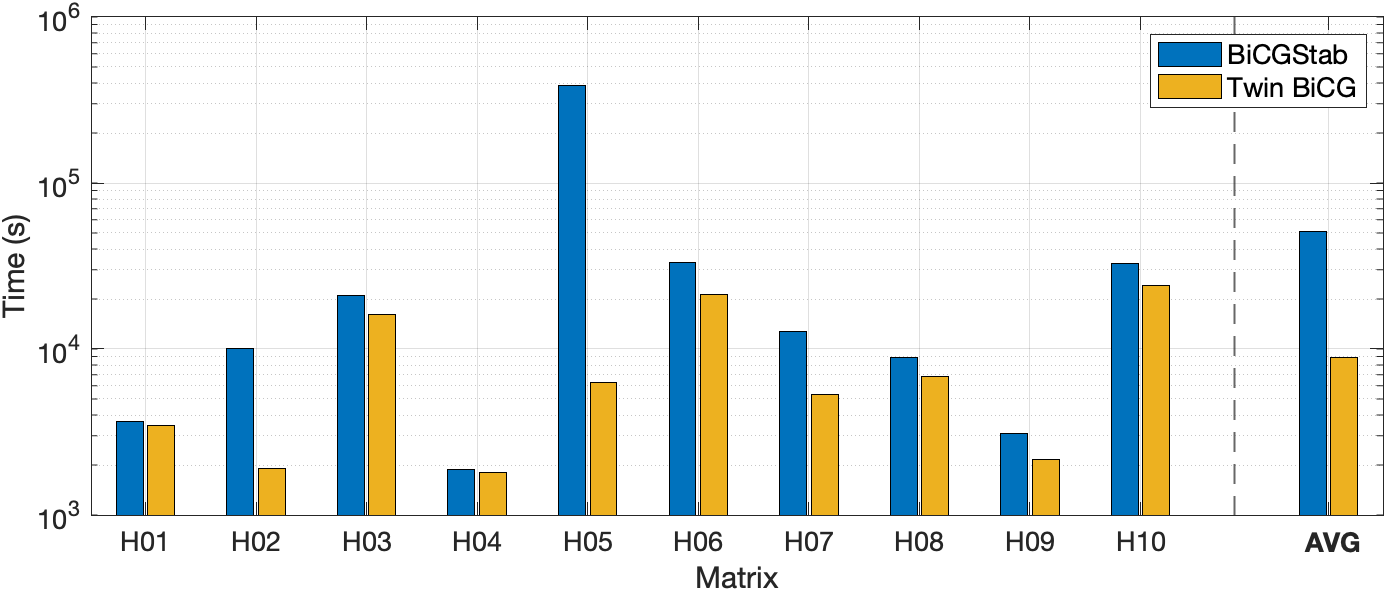} 
   \caption{Trace times comparing BiCGStab and Twin BiCG on a single computational thread for the H matrices ($12^3\times 16$). The relative solve tolerance is $10^{-3}$ and the trace tolerance is $12^3\times 16\times 0.001$.}
   \label{fig:12121216_singleThread}
\end{figure}

The test results are displayed in Fig.~\ref{fig:12121216_singleThread}. For matrices H02, H05, H06, and H07, Twin BiCG was substantially faster because several BiCGStab solves failed to converge, which significantly increased the total time. Twin BiCG avoids this since $\pi$ is reused from a successful solve. For matrices H01, H04, H09, and H10, both methods finished in comparable and relatively short times. Twin BiCG often requires more iterations per solve, but each iteration is cheaper because it needs fewer MVPs and no dot products beyond the first solve, so the two effects roughly offset. Twin BiCG is therefore competitive with BiCGStab when BiCGStab converges and substantially more robust when it does not.

\subsection{A matrices}\label{sec:Amats}
\begin{figure}[h] 
   \centering
   \includegraphics[width = \figWidth in]{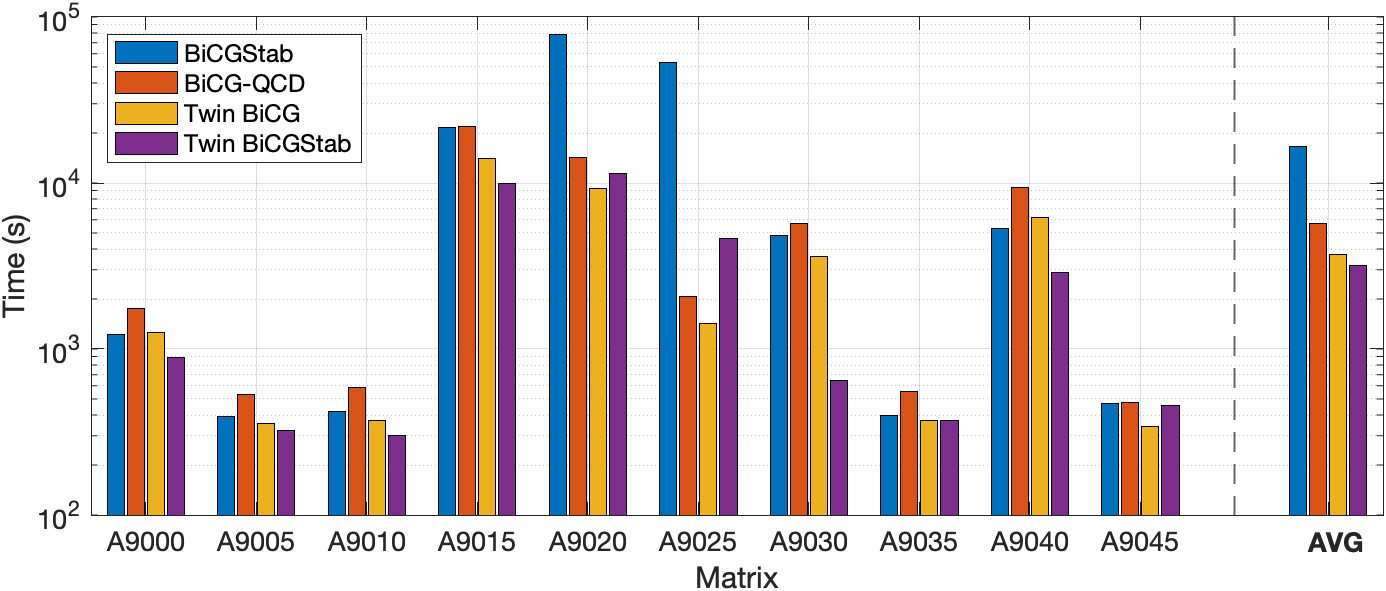} 
   \caption{Trace times for the A matrices ($16^3\times 48$) using BiCGStab, BiCG-QCD, Twin BiCG, and Twin BiCGStab. The relative solve tolerance is $10^{-3}$ and the trace tolerance is $0.001\times 16^3\times 48$.}
   \label{fig:A_mphys}
\end{figure}

Now we present GPU tests on the A matrices. We compare four methods: Twin BiCG, Twin BiCGStab, BiCGStab and BiCG-QCD, where the last two methods serve as benchmarks. The times required to compute the trace on a set of ten different matrices are presented in Fig.~\ref{fig:A_mphys}. 

BiCGStab fails on 54\% and 86\% of noise vectors for matrices A9020 and A9025, respectively, compared to 0\% for all other matrices in the test set. This difficulty may be attributed to the presence of negative eigenvalues in their spectra, while the spectra of the other matrices lie in the right-half plane.\footnote{The spectra were computed using the polynomial-preconditioned Arnoldi algorithm.} Twin BiCGStab circumvents this problem by reusing a successful $\pi$ polynomial on noise vectors that would otherwise fail to converge under standard BiCGStab. As a result, it was faster than BiCGStab by factors of 6.8 and 11.5 on matrices A9020 and A9025, respectively. %Twin BiCGStab is therefore more robust and faster than standard BiCGStab on these matrices.

Matrices A9020, A9025, and A9045 are the only instances where Twin BiCGStab is slower than Twin BiCG. For A9025, this is partly due to the first 15 noise vectors, all of which failed to converge. Until a successful solve occurs, there is no $\pi$ to reuse, so each noise vector is a full solve attempt. Many of these attempts ran the maximum number of solve iterations before terminating, and each iteration of Twin BiCGStab costs two MVPs, so they contribute a significant share of the total trace time. 

The iteration count associated with each solve also matters. Let $r$ be the ratio of the number of iterations per noise vector for Twin BiCGStab to that for Twin BiCG. For A9020, A9025, and A9045, $r = 0.93$, 1.32, and 0.76, respectively. For all other matrices $r$ averaged 0.48---roughly half as many iterations for Twin BiCGStab compared to Twin BiCG. Because each iteration of Twin BiCGStab requires two MVPs whereas Twin BiCG only requires one, Twin BiCGStab is only faster when $r$ is sufficiently small; the relatively large values of $r$ on these matrices help explain the larger times. In short, although Twin BiCGStab is generally faster and more robust that BiCGStab, its cost is dominated by the number of iterations needed for the first successful solve, which is determined by the underlying BiCGStab algorithm.

The average times across the matrices for each method are shown on the right of Fig.~\ref{fig:A_mphys}. Because estimating the trace across an entire ensemble requries solving on every configuration, this average is the most important metric. On this measure, Twin BiCGStab is the fastest method and standard BiCGStab is the slowest, which demonstrates the benefit of recycling the $\pi$ polynomial.

In addition to the runs in Fig.~\ref{fig:A_mphys}, tests were also done in a batched format. The total number of noise vectors required for the trace is not known in advance, but because the error scales as $1/\sqrt{n}$, one may estimate the number of noise vectors from the first batch, and the next batches are sized accordingly. (However, the size of the first batch is an \textit{a priori} heuristic guess.) The benefit of batching is that serial matrix-vector operations are replaced with matrix-matrix operations which reuse data more efficiently, reducing the trace time without compromising the accuracy of the trace estimate.

\begin{figure}[h] 
   \centering
   \includegraphics[width = \figWidth in]{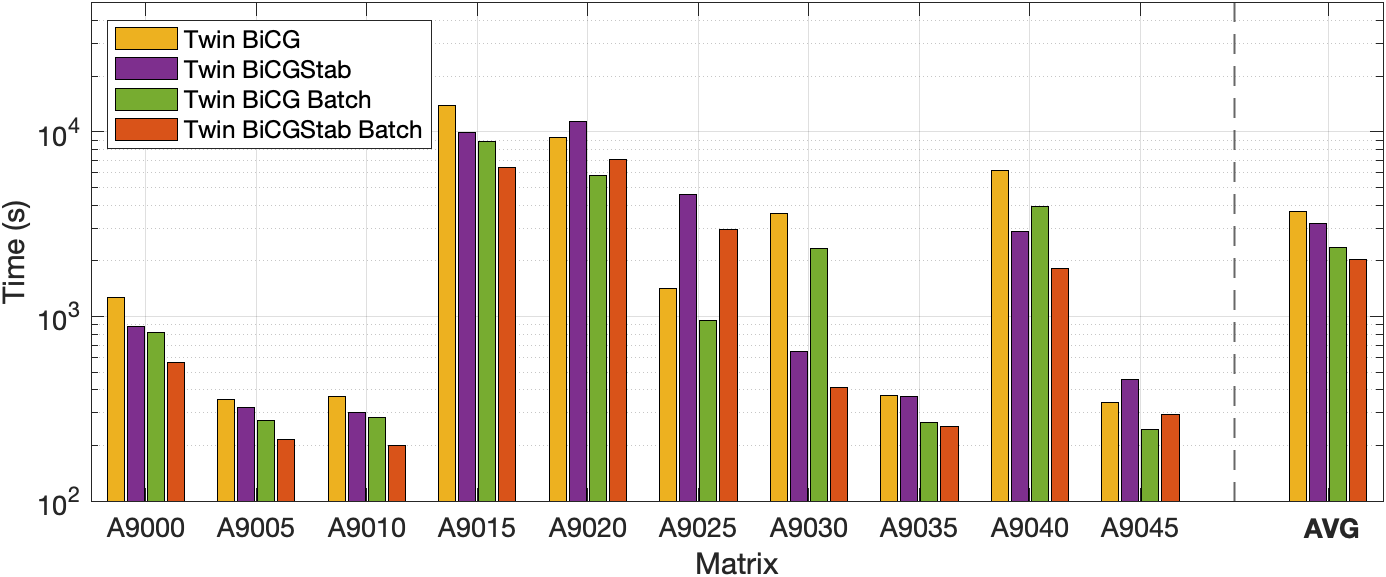} 
   \caption{Trace times for the A matrices ($16^3\times 48$) using Twin BiCG, Twin BiCGStab, Twin BiCG Batch, and Twin BiCGStab Batch. The relative solve tolerance is $10^{-3}$ and the trace tolerance is $0.001\times 16^3\times 48$.}
   \label{fig:A_mphys_batched}
\end{figure}

The results in Fig.~\ref{fig:A_mphys_batched} include batched methods. They are consistent with the relative ranking of the results in Fig.~\ref{fig:A_mphys}. In every test, batching of a method made it faster than its non-batched counterpart. This is expected since batching just changes how the operations are executed, not the algorithm structure. These results suggest that batching is always preferable to serial solves when memory permits. Table~\ref{tbl:speedup} gives the speedup of each method relative to the slowest, standard BiCGStab.

\subsection{B matrices}\label{sec:Bmats}
\begin{figure}[h] 
   \centering
   \includegraphics[width = \figWidth in]{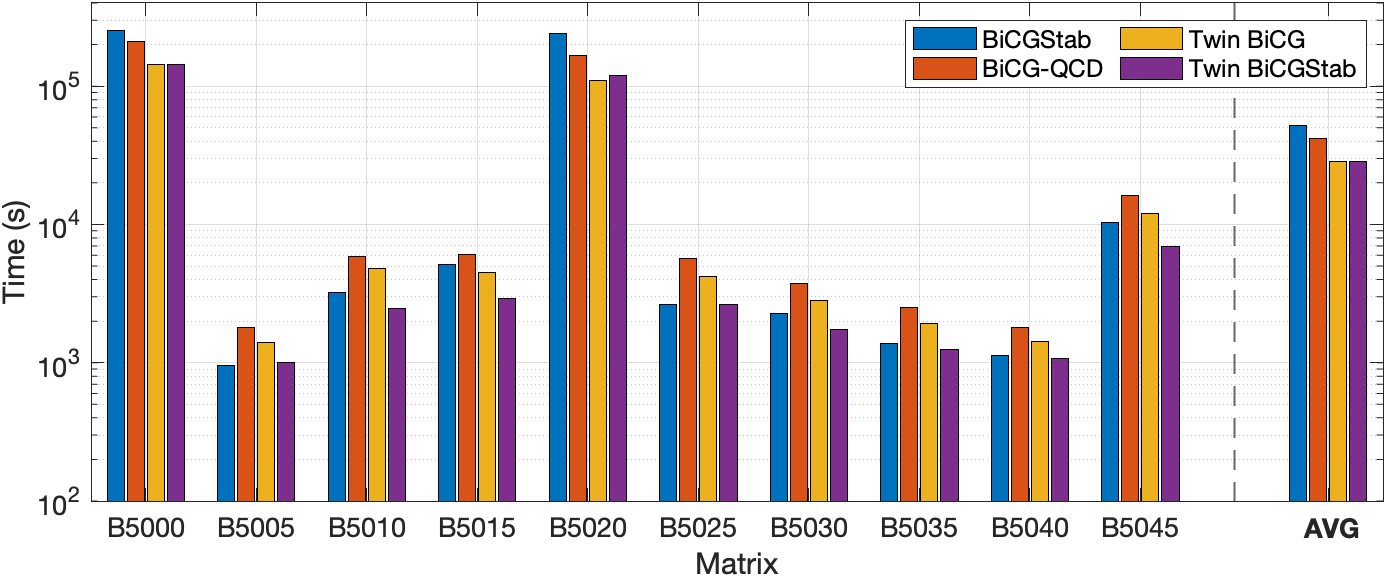} 
   \caption{Trace times for the B matrices ($24^3\times 64$) using BiCGStab, BiCG-QCD, Twin BiCG, and Twin BiCGStab. The relative solve tolerance is $10^{-3}$ and the trace tolerance is $0.0005\times 24^3\times 64$.}
   \label{fig:B_mphys}
\end{figure}

We now present GPU tests on the larger B matrices. The results are shown in Fig.~\ref{fig:B_mphys}. Unlike the set of A matrices, BiCGStab was faster than both BiCG-QCD and Twin BiCG on most B matrices. The exceptions were B5000 and B5020, on which BiCGStab failed to converge on 14.3\% and 28.9\% of noise vectors, respectively. These failures were costly enough that BiCGStab had the longest average wall-clock time of all the methods. As with the A matrices, Twin BiCG and Twin BiCGStab are the fastest average methods because they avoid repeated failures by reusing a successful $\pi$ polynomial rather than solving each noise vector independently.

\begin{table}[]
\centering
\begin{tabular}{lllllll} \hline\hline
Solver              & A speedup & B speedup \\ \hline
BiCGStab            & 1.00      & 1.00      \\
Twin BiCG           & 4.47      & 1.82      \\
Twin BiCGStab       & 5.22      & 1.84      \\
BiCG-QCD            & 2.91      & 1.25      \\
Twin BiCG Batch     & 7.00      &           \\
Twin BiCGStab Batch & 8.21      &          \\ \hline\hline
\end{tabular}
\caption{Speedup factor for Hutchinson's method using different solvers compared to the slowest average solver, BiCGStab. Batched tests were not performed for B matrices.}
\label{tbl:speedup}
\end{table}

\section{Singular Value Deflation and Polynomial Subtraction} \label{sec:PSandSVD}
The costs of the trace estimation can be reduced by subtracting a polynomial approximation of $A^{-1}$~\cite{MultPolyMonteCarlo}. The trace may be written as
\begin{equation}
    \mathrm{Tr}(A^{-1}) = \mathrm{Tr}(A^{-1} - p(A)) + \mathrm{Tr}(p(A)),
    \label{eq:tracePS}
\end{equation}
where $p(A) \approx A^{-1}$.\footnote{$p(A)$ is computed using the Nonsymmetric Lanczos algorithm on $A$ (special for QCD).} The degree of the polynomial must be high so that very few noise vectors are required to compute the first trace in the sum, $\mathrm{Tr}(A^{-1}-p(A))$. Costs can be further reduced by including a singular value deflation in Eq.(\ref{eq:tracePS}).  The idea of deflating eigenvalues and singular values from a QCD Monte Carlo started with~\cite{QCDsubtr}, followed by~\cite{QCDsubtrDiss} in which singular value deflation is called “Hermitian forced eigenvalue subtraction” (but there are some errors). 
The full trace is
\begin{equation}
\mathrm{Tr}(A^{-1}) \approx \mathrm{Tr}(A^{-1}-p(A)) + \mathrm{Tr}\left(p(A) - \sum_i \frac{1}{\sigma_i}v_i u_i^\dagger\right) + \mathrm{Tr}\left(\sum_i \frac{1}{\sigma_i}v_i u_i^\dagger\right),
\label{eq:fullTrace}
\end{equation}
where $\{1/\sigma_i\}$ are the smallest singular values of $A^{-1}$, and $\{v_i\}$ and $\{u_i\}$ are the left and right singular values of $A^{-1}$, respectively.\footnote{The singular values and vectors are obtained through the symmetric Lanczos algorithm on $\gamma_5 A$.} The number of noise vectors needed for the second trace in Eq.(\ref{eq:fullTrace}) will be reduced from the number required for the second trace in Eq.(\ref{eq:tracePS}) since the problematic singular modes are deflated out. Trace three of Eq.(\ref{eq:fullTrace}) can be computed exactly, so it will not contribute any error in the estimation.

\subsection{Testing}
\begin{figure}[h] 
   \centering
   \includegraphics[width = \figWidth in]{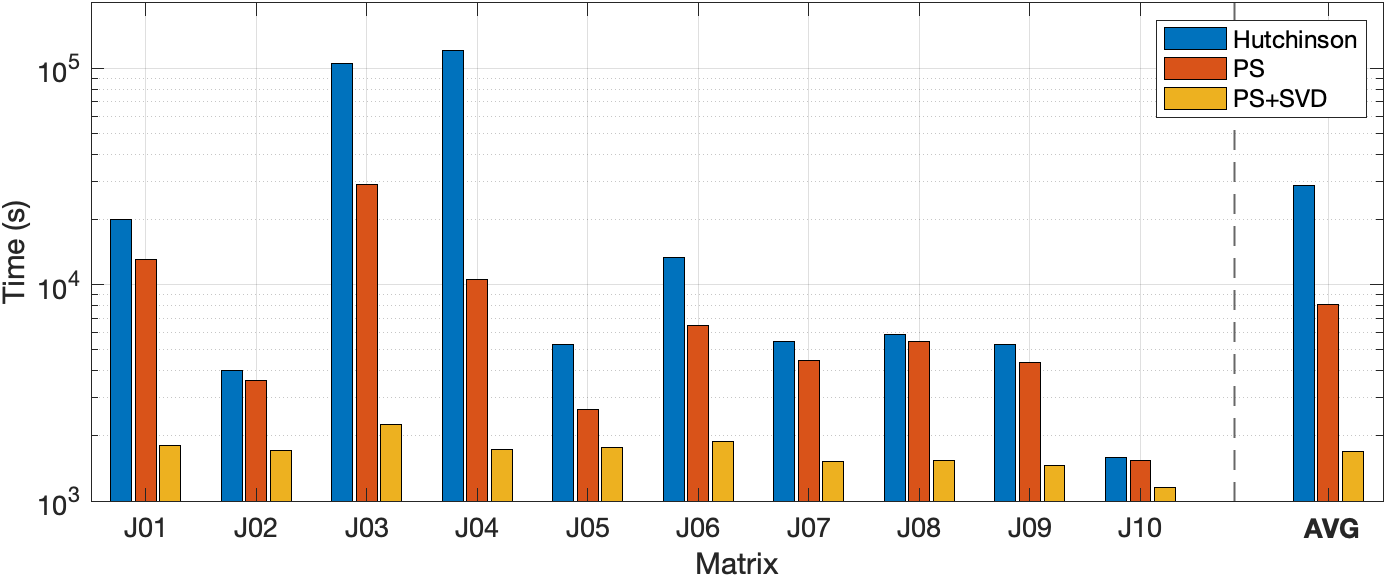} 
   \caption{Trace times for the J matrices ($24^4$) using Hutchinson's method, the PS trace, and the PS+SVD trace. Traditional Hutchinson's method is solved using BiCGStab with a relative solve tolerance of $10^{-5}$. The trace tolerance is $0.0005\times 24^4$.}
   \label{fig:SVD}
\end{figure}

The polynomial subtracted (PS) trace (Eq.(\ref{eq:tracePS})) and the full polynomial subtracted and singular value deflated (PS+SVD) trace (Eq.(\ref{eq:fullTrace})) were tested on a $24^4$ quenched lattice with Wilson fermions at the critical hopping parameter, $\kappa = \kappa_c = 0.157$, using NVIDIA A100 GPUs. The fermion matrices associated with this lattice are labeled with a ``J'' and the wall-clock times to compute the trace using the PS and PS+SVD methods are shown in Fig.~\ref{fig:SVD}. These times are compared to standard Hutchinson's method using BiCGStab.

On average, PS alone is 3.52 times faster than Hutchinson's method, and adding the deflation raises the speedup to 17.0 (see Table~\ref{tbl:SVDspeedup}). The individual matrices show how PS and SVD contribute different effects. PS reduces the time by a roughly constant factor, so matrices that are slow under Hutchinson are still relatively slow under PS. Deflation, by removing the smallest singular values that make the solves difficult, makes the times similar across the matrices.% This is reflected in the coefficients of variation\footnote{The coefficient of variation is defined as the standard deviation divided by the average: $\sigma/\mu$.} of $1.56$, $1.01$, and $0.111$ for the Hutchinson, PS, and PS+SVD traces, respectively. The much smaller relative spread in PS+SVD indicates that deflation has an equalizing effect and makes the cost far less dependent on the difficulty of the individual matrix.

\begin{table}[]
\centering
\begin{tabular}{llll} \hline\hline
             & Hutch/PS & PS/(SVD+PS) & Hutch/(SVD+PS) \\ \hline
Speedup factor & 3.52   &  4.82       & 17.01  \\ \hline\hline
\end{tabular}
\caption{Speedup factor when comparing Hutchinson's method (Hutch), the single polynomial subtraction method of Eq.(\ref{eq:tracePS}) (PS), and the full polynomial subtracted and singular value deflated method of Eq.(\ref{eq:fullTrace}) (SVD+PS).}
\label{tbl:SVDspeedup}
\end{table}

Computing the singular values used in the deflation accounts for only 11\% of the overall PS+SVD trace time on average, which is included in the times reported in Fig.~\ref{fig:SVD}. The number of singular values deflated was not fixed in advance and cannot be easily predicted. This does not appear to be a practical obstacle provided enough of the singular modes are deflated to reduce the overall difficulty of the problem.

\section{Conclusions}
We first introduced a new class of Krylov solvers called Twin methods for solving multiple RHSs. We then used Twin BiCG and Twin BiCGStab to compute $\mathrm{Tr}(A^{-1})$ with Hutchinson's method on several different semi-quenched lattice QCD matrices from two MILC ensembles. Ensemble A was a $16^3\times 48$ lattice and B was a $24^3 \times 64$ lattice; the kappa values associated with the physical pion mass were used. Twin BiCG and Twin BiCGStab were especially beneficial on matrices where standard solvers failed on a large fraction of noise vectors, since the Twin methods reuse the $\pi$ polynomial from a successful solve and so avoid repeated failures. Batching noises further improved the trace times because matrix-matrix operations reuse data more efficiently than matrix-vector operations. On the A matrices, Twin BiCGStab was 8.21 times faster than the slowest method, standard BiCGStab. We then introduced a polynomial subtraction and a singular value deflation to the trace. For these tests we used a $24^4$ quenched lattice set to $\kappa$-critical. The PS+SVD method gave a 17-fold speedup over standard Hutchinson's method and made the cost far less dependent on the individual matrix.

\section*{Acknowledgements}
We are grateful to the MILC collaboration for sharing the lattices used in this work. This work was supported in part through allocation MTH250023 from the NSF ACCESS program.

%\begin{thebibliography}{99}
\bibliographystyle{unsrt}
\bibliography{bibliography_BL_Lattice2026.bib}

@article{FROMMER_1995,
   title={MANY MASSES ON ONE STROKE: ECONOMIC COMPUTATION OF QUARK PROPAGATORS},
   volume={06},
   ISSN={1793-6586},
   url={http://dx.doi.org/10.1142/S0129183195000538},
   DOI={10.1142/s0129183195000538},
   number={05},
   journal={International Journal of Modern Physics C},
   publisher={World Scientific Pub Co Pte Lt},
   author={Frommer, Andreas and N\"{o}ckel, Bertold and G\"{u}sken, Stephan and Lippert, Thomas and Schilling, Klaus},
   year={1995},
   month=Oct, pages={627–638} }

@article{HutchTrace,
  Author={M. F. Hutchinson},
  Title="A stochastic estimator of the trace of the influence matrix for {L}aplacian smoothing splines",
  Journal={Commun. Stat. Simul. Comput.},
  Year=1989,
  Volume=18,
  Pages={1059-1076}     }

@article{BiCGStab,
  Author={H. A. van der Vorst},
  Title="{B}i-{CGSTAB}: A Fast and Smoothly Converging Variant
of {B}i-{CG} for the Solution of Nonsymmetric Linear Systems",
  Journal={SIAM J. Sci. Stat. Comput.},
  Year={1992},
  Volume=13,
  Pages={631-644}     }

@article{PhysRevD.87.054505,
  title = {Lattice QCD ensembles with four flavors of highly improved staggered quarks},
  author = {Bazavov, A. and Bernard, C. and Komijani, J. and DeTar, C. and Levkova, L. and Freeman, W. and Gottlieb, Steven and Zhou, Ran and Heller, U. M. and Hetrick, J. E. and Laiho, J. and Osborn, J. and Sugar, R. L. and Toussaint, D. and Van de Water, R. S.},
  collaboration = {MILC Collaboration},
  journal = {Phys. Rev. D},
  volume = {87},
  issue = {5},
  pages = {054505},
  numpages = {19},
  year = {2013},
  month = {Mar},
  publisher = {American Physical Society},
  doi = {10.1103/PhysRevD.87.054505},
  url = {https://link.aps.org/doi/10.1103/PhysRevD.87.054505}
}

@article{MultPolyMonteCarlo,
  Author={P. Lashomb and R. B. Morgan and T. Whyte and W. Wilcox},
  Title="Multi-polynomial {M}onte {C}arlo for {QCD} Trace Estimation",
  Journal={Comput. Phys. Commun.},
  Volume={300},
  pages={109163-1 - 109163-10},
  year={2024}}

@article{QCDsubtr,
  Author={V. Guerrero and R. B. Morgan and W. Wilcox},
  Title="Eigenspectrum Noise Subtraction Methods in Lattice {QCD}",
  Journal={Proceedings of Science, LAT2009},
  Year=2009,
  Pages={041/1-041/8}  }

@unpublished{QCDsubtrDiss,
  Author={V. Guerrero },
  Title="Electric Neutron Polarizability and Eigenspectrum Subtraction Techniques for Disconnected Quark Loops",
  Year=2011,
  Note = "PhD Thesis, Baylor University"  }
%\bibliography{MORGAN}

% \end{thebibliography}

\end{document}